\documentclass[
  a4paper,
  twocolumn,
  longbibliography,
  aps,
  pra,
  superscriptaddress,
  amsfonts,amsmath,amssymb,
  floatfix
]{revtex4-2}
\usepackage{changes}
\usepackage[
  colorlinks,
  citecolor=blue,
  linkcolor=blue,
  urlcolor=blue
]{hyperref}
\usepackage{graphicx,epstopdf, bm}
\usepackage{braket}
\usepackage{blindtext}
\usepackage{enumitem}
\usepackage[capitalize]{cleveref}
\usepackage{xcolor}
\usepackage[all]{hypcap}
\usepackage{cleveref}
\usepackage{amssymb}

\crefname{section}{Sec.}{Secs.}
\Crefname{section}{Section}{Sections}
\crefname{appendix}{App.}{Appces.}
\Crefname{appendix}{Appendix}{Appendices}
\usepackage{hyperref}
\usepackage{xcolor}

\begin{document}

\title{{Enhancing optomechanical force sensing utilizing synthetic magnetism}}


\author{Ding-hui Xu, Zheng Liu, and Chang-shui Yu}
	\email{Electronic address: ycs@dlut.edu.cn}
	\address{School of Physics, Dalian University of Technology, Dalian 116024, P.R. China}
\date{\today}


\begin{abstract}
{In precision force sensing of multi-mechanical mode optomechanical systems, coherent interference can decouple certain degenerate vibrational modes from the cavity field, leading to incomplete information regarding the measured signal. In this paper, we propose a scheme to enhance and control the detection bandwidth in optomechanical force sensing by exploiting synthetic magnetism achieved through tuning phonon hopping interactions. By toggling between broken and unbroken dark mode, this approach effectively manages the response bandwidth and exhibits intriguing additional noise characteristics. Specifically, when the dark mode remains unbroken, the thermal noise is robust and reduced to half of that of a standard device. In contrast, when the dark mode is broken, thermal noise increases substantially at mechanical resonance but remains the same as when the dark mode is unbroken at effective detection frequencies. Moreover, our scheme offers the dual benefit of amplifying the mechanical response while suppressing additional noise, with the potential to surpass the standard quantum limit.}

\end{abstract}


\maketitle

\section{Introduction}
  \label{sec:introduction}

 In recent years, quantum-enhanced sensing has capitalized on the specific advantages of quantum resources, achieving significant success on various platforms \cite{doi:10.1126/sciadv.adg1760, Chen2023, Backes2021, doi:10.1126/science.abi5226}.
A notable example is the enhancement of gravitational wave detection in LIGO, which is enabled by quantum correlations between light and kilogram-scale mirrors \cite{galaxies10010036}. The principle of squeezed light has been widely adopted in quantum sensing technologies, including the development of quantum magnetometers with entangled twin beams capable of functioning in challenging environments \cite{PedrozoPenafiel2020}. Moreover, recent progress in the characterization of non-Gaussian entangled states of superconducting qubits has led to improved measurement precision \cite{PhysRevLett.128.150501}. These advancements highlight substantial breakthroughs in the realm of quantum sensing. 

Optomechanical cavities (COM) offer a powerful platform for high-precision sensing by enabling optical detection of minute mechanical displacements \cite{RevModPhys.86.1391, PhysRevLett.108.120801,li2021cavity}.  Multiple fields of high-performance sensors have been proposed in both experimental and theoretical studies, including nanoscale optomechanical pressure sensors based on ring resonators on thin membranes \cite{Zhao:12}, nanomechanical displacement sensors \cite{liu2020integrated}, and acoustic sensors utilizing optical resonances \cite{Yao:20}. However, when precision measurements are performed using quantum techniques, Heisenberg's uncertainty principle \cite{BUSCH2007155} imposes a fundamental lower limit on the sensitivity of the system, known as the standard quantum limit (SQL) \cite{doi:10.1126/science.1104149}. Therefore, many weak force sensing schemes aim to reduce quantum noise, such as introducing auxiliary mechanical oscillators \cite{PhysRevA.99.063811,doi:10.1073/pnas.1608412114,liuweak,PhysRevLett.127.073601},  using squeezed optomechanics \cite{wang2024quantum,zhao2020weak, Zhang:24,zhang2024quantum}, employing feedback control \cite{PhysRevApplied.17.034020, PhysRevA.98.023828, PhysRevApplied.19.054091}, and performing ground-state cooling and quantum-state control of mechanical oscillators \cite{chan2011laser} and so on. \begin{figure}[b]
        \centering
        \includegraphics[width=\linewidth]{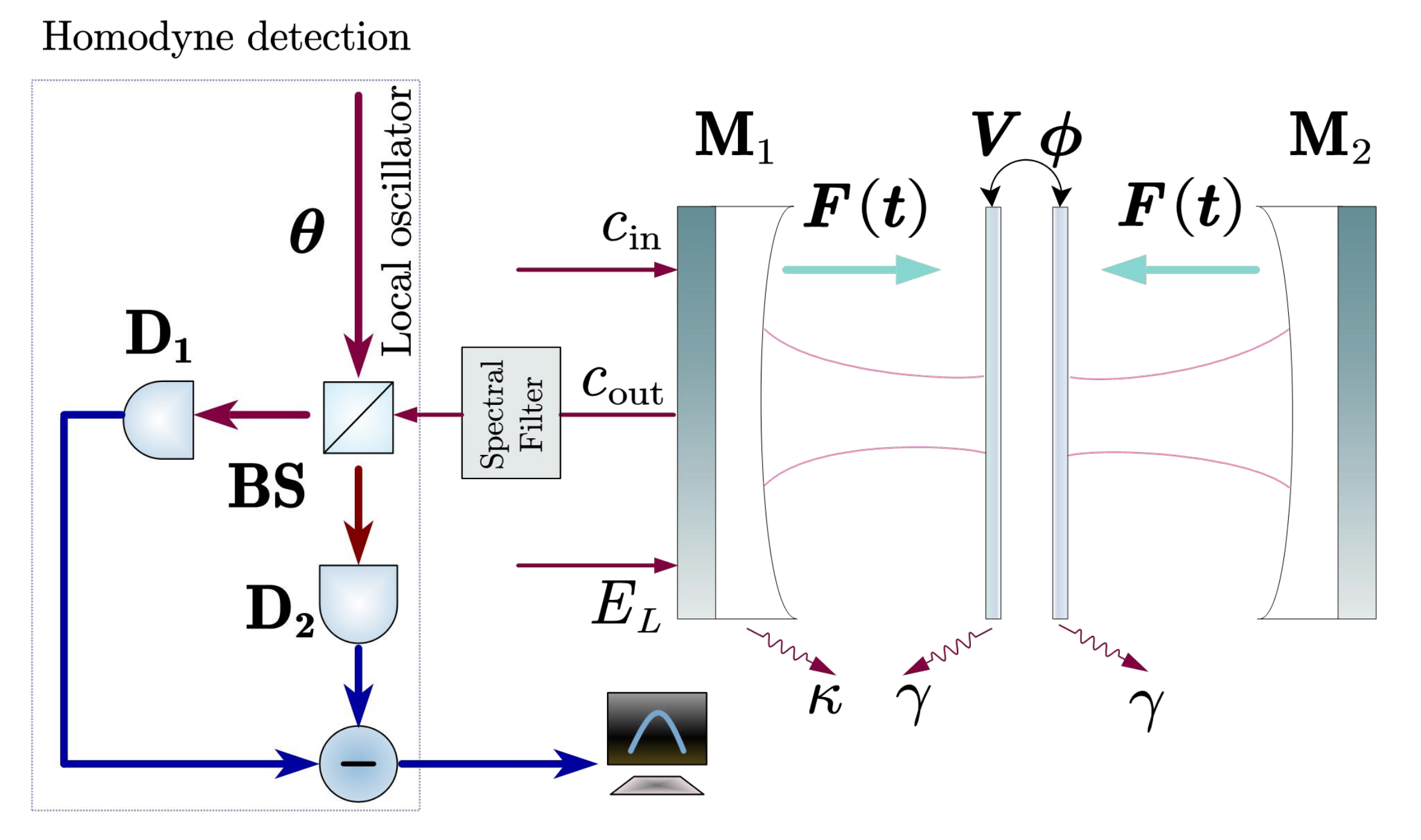}
        \caption{The schematic diagram of the optomechanical force sensor model. The system consists of an optical cavity and two mechanically coupled oscillators, with their coupling dependent on the phase. The coupling strength, denoted by $V$, and the coupling phase, denoted by $\phi$, serve as probes for detecting ideal impulsive forces. The right-end mirror is a perfect cavity mirror, free of dissipation, while the transmitted light at the left-end mirror passes through an amplitude spectrum filter. The output from the homodyne detection setup can detect external force signals. The homodyne detection setup includes a beam splitter, photodiode, local oscillator light with a phase $\theta$, and a subtractor.
}
        \label{fig:1}
 \end{figure}
 
Recently, multimode optomechanical systems involving two or more mechanical oscillators \cite{doi:10.1126/science.abf5389, Yang2020, Spethmann2016} have made progress in cavity optomechanics. It has been observed that, due to coherent interference, certain degenerate vibrational mechanical modes become coupled to the optical field, making them undetectable by the optical field and giving rise to dark mode effects \cite{dong2012optomechanical, PhysRevA.96.023860, PhysRevA.102.011502, PhysRevLett.108.153604}. For example, dark modes hinder the ability to achieve ground-state cooling of multiple oscillators \cite{PhysRevA.99.023826}, destroy quantum entanglement \cite{PhysRevLett.126.163604}, thus presenting a significant challenge in physical research. To break the dark modes (by modulating the photon-phonon hopping rates), {several theoretical and experimental schemes have been proposed using synthetic magnetism} \cite{Massembele2024, fang2017generalized, xu2015optical, Schmidt:15}. It has been shown that synthetic magnetism in optomechanical systems can enhance entanglement generation \cite{PhysRevLett.129.063602, Massembele2024}, control soliton waves in mechanical arrays \cite{ALPHONSE2022111593}, enable nonreciprocal optical behavior in silicon optomechanical circuits \cite{fang2017generalized}, and realize photon transport via synthetic magnetic field design \cite{Wang_2020}. However, leveraging dark mode effects can enhance various other aspects of cavity optomechanics \cite{doi:10.1126/sciadv.1501142, lai2020tunable, PhysRevB.101.045409}.
  
  Due to the intriguing physical properties introduced by synthetic magnetism, we utilize synthetic magnetism to enhance  optomechanical weak force sensing. Our approach involves tuning the coupling strength $V$ and phase $\phi$ between two mechanical resonators modes to form a phonon-hopping interaction, which induces a synthetic gauge field \cite{doi:10.1126/science.aay3183,celi2014synthetic, PhysRevLett.113.087403}. This, in turn, hybridizes into bright and dark modes that decouple from the system. By adjusting the coupling phase, we find that multimode quantum devices can flexibly switch between high- and low-frequency regimes, significantly overcoming the narrow bandwidth limitations imposed by mechanical resonances and revealing the potential to surpass the SQL.
  {The remainder of the paper is organized as follows. Section~\ref{sec:Model} provides a brief introduction to the model for the optomechanical force sensing system. By solving the dynamical equations, we ultimately derive the weak-field sensitive detection spectrum density. In Section~\ref{sec: results}, we analyze the physical sensing properties related to reducing quantum noise and increasing detection bandwidth. The experimental feasibility is discussed in Section~\ref{sec:experiments}, and the conclusion is presented in Section~\ref{sec:conclusion}.}


\section{The model and the dynamics}
  \label{sec:Model}
The setup of our weak force sensing is illustrated in Fig. \ref{fig:1}. Two identical nanomembranes are placed inside a high-quality Fabry-Pérot cavity driven by an external laser. Phonons can directly couple with photons through the radiation pressure of the optical cavity field. The minute displacements of these oscillators can be reflected in the phase shift of the cavity field. Combined with a standard homodyne detection setup, this system can serve as an ultrasensitive microforce detection device. Our scheme considers the phase-correlated phonon hopping interaction between the two mechanical membranes.  The Hamiltonian of our system in a rotating frame with the laser driving frequency $\omega_L $  reads{
\begin{align}
\hat{H}& = \hbar \Delta \hat{c}^\dagger \hat{c} +\hbar  \displaystyle\sum\limits_{i=1,2} (\omega_m\hat{b}_i^\dagger \hat{b}_i\ +  g c^\dagger \hat c(\hat b_i +\hat  b_i ^\dagger) \nonumber \\
&\quad+\hbar V ( e^{i\phi}\hat b_1^\dagger\hat b_2 + e^{-i\phi}\hat b_1 \hat b_2^\dagger )
 -x_{\text{ZPF}} F(t)\displaystyle\sum\limits_{i=1,2}(\hat{b}_i + \hat{b}_i^\dagger)\nonumber \\
&\quad+i\hbar E_L (\hat{c}^\dagger - \hat{c}), \label{E1} 
\end{align}}where $\Delta = \omega_c - \omega_L$ is the laser detuning of the cavity mode,  {\( \hat{c}^\dagger \) (\(\hat{ c} \)) and \(\hat{ b}_j^\dagger \) (\( \hat{b}_j \)) are the creation (annihilation) operators of the cavity-field mode (with resonance frequency \( \omega_c \)) and the \( i \)-th vibrational mode (with resonance frequency \( \omega_m \)).} The first two terms  represent the free Hamiltonian of the cavity field and the two mechanical oscillators. The zero-point position is given by  $x_{\text{ZPF}} = \sqrt{ \frac{\hbar}{2m\omega_m} }$ and the single-photon optomechanical-coupling strength is $g =  \sqrt{2}(\frac{\partial \omega_c}{\partial x}) x_{\text{ZPF}}$ .
The fourth term of the Hamiltonian depicts the phase-dependent phonon-hopping interaction, which can induce a reconfigurable synthetic gauge field. {In classical coupled mechanical oscillators, the modes generally exhibit a fixed phase difference \cite{doi:10.1126/science.abf5389}. However, We propose phase-dependent phonon hopping interactions in a one-dimensional (1D) optomechanical crystal system realized by pumping the optomechanical cavities with phase-correlated lasers \cite{fang2017generalized,PhysRevLett.129.063602}.  The phase difference is controlled using a fiber-optic phase shifter, enabling the realization of a stable synthetic magnetic field
  \cite{Habraken_2012}. Ref. \cite{PhysRevLett.129.063602}  has provided a detailed derivation of the terms associated with synthetic magnetism. Additionally, synthetic magnetism can also be achieved by coupling two vibrational modes to a superconducting charge qubit in circuit-mechanical systems \cite{Massel2012}}. $E_L = \sqrt{ \frac{P_L \kappa_{\text{in}}}{\hbar \omega_L} }$ is the driving strength of the external laser field, where $P_L$ represents the input power of the coherent driving field. {The input classical force $F(t) $= $  \sqrt{\frac{m\omega_m}{\hbar}}F_{\text{ext}}  $ is  imposed on the mechanical membrane by the coupling of the mechanical membrane in the horizontal direction.} 
  
 {By defining the optical quadrature operators $\hat{X}_i = \frac{(\hat{o}_i^\dagger + \hat o_i)}{\sqrt{2}}  $ and $\hat{P}_i = \frac{i(\hat o_i^\dagger - \hat o_i)}{\sqrt{2}}$, where \(\hat o_i \) and \( \hat {o}_i^\dagger \) are the annihilation and creation operators for vibrational mode, respectively,
the Hamiltonian becomes} {
\begin{align}
\hat{H}& = \hbar \Delta \hat{c}^\dagger \hat{c} + \frac{\hbar \omega_m}{2} \sum_{i=1,2} ( \hat{P}_i^2 + \hat{X}_i^2 ) + \hbar g \hat{c}^\dag \hat{c} (\hat{X}_1 + \hat{X}_2) \nonumber \\
&  + \hbar V [(\cos{\phi} \hat{X}_1 +\sin{\phi} \hat{P}_1)\hat{X}_2 +(\cos{\phi} \hat{P}_1- \sin{\phi} \hat{X}_1) \hat{P}_2]\nonumber\\
&  + i \hbar E_L (\hat{c}^\dagger - \hat{c}) - F_{\text{ext}}(\hat{X}_1 + \hat{X}_2).\label{E2} 
\end{align}}The system's dynamics subject to its external environments can be described by the quantum Langevin equation \cite{PhysRevLett.46.1} as 
\begin{eqnarray}\label{E3}
 \dot{\hat{c}} &=& -\left( i\Delta + \kappa \right) \hat{c} - i g \hat{c} (\hat{X}_1 + \hat{X}_2) + E_L + \sqrt{2 \kappa} \hat{c}_{in} , \notag \\
 \dot{\hat{X}}_1& =& \omega_m \hat{P}_1 + V \sin{\phi} \hat{X}_2 + V \cos{\phi} \hat{P}_2, \notag \\
\dot{\hat{X}}_2 &=& \omega_m \hat{P}_2 - V \sin{\phi} \hat{X}_1 + V \cos{\phi} \hat{P}_1, \notag \\
\dot{\hat{P}}_1 &=& -\omega_m \hat{X}_1 - g \hat{c}^\dag\hat{c}  - V \cos{\phi} \hat{X}_2 + V \sin{\phi} \hat{P}_2 \notag \\
& - &\gamma \hat{P}_1 + \sqrt{2 \gamma} {f}_{in,1}, \notag \\
 \dot{\hat{P}}_2 &=& -\omega_m \hat{X}_2 - g \hat{c}^\dag\hat{c}  - V \cos{\phi} \hat{X}_1 - V \sin{\phi} \hat{P}_2 \notag \\
& - &\gamma \hat{P}_2 + \sqrt{2 \gamma} {f}_{in,2},
\end{eqnarray}
where the operator $ \hat c_{\text{in}} $ represents the cavity field input noise with the correlation function  $ \langle \hat c_{\text{in}}(t) \hat c_{\text{in}}^\dagger(t') \rangle = \delta(t - t') $, $ \kappa $ is the decay rate of the cavity, and $\gamma $ is the damping rate of the nanomechanical membrane.
In Eq. (\ref{E3}), \( f_{\text{in}, i} = f_{\text{ext}} + \hat{f}_{\text{th}, i} \), denotes the input mechanical operator, consisting of the force to be detected \( f_{\text{ext}} = \sqrt{\frac{1}{2m\hbar\gamma \omega_m}} F(t) \) and the thermal noise operator \( \hat{f}_{\text{th}, i} = \sqrt{\frac{1}{2m\hbar\gamma_m \omega_m}} \hat{F}_{\text{th}, i} \), with the correlation function \( \langle \hat{F}_{\text{th}, i}(t) \hat{F}_{\text{th}, i}(t') \rangle \approx \hbar m \gamma \omega_m (2 \bar{n} + 1) \delta(t - t') \) \cite{PhysRevA.63.023812}, where \( \bar{n} = \left( \exp\left(\frac{\hbar \omega_m}{k_B T}\right) - 1 \right)^{-1} \) represents the thermal occupation number for a given temperature \( T \).

With the strong driving and optomechanical weak coupling condition, the quantum operators in the Langevin equations can be expressed into their semi-classical values and the small one-order fluctuations,  i.e., $\hat{c} = \bar{c} + \delta\hat{c}$, $\hat X_i = \bar{X}_i + \delta\hat{X}_i$, $\hat P_i = \bar{P}_i + \delta\hat{P}_i$. In this way, the above Langevin equations can be rewritten into two groups of equations, one of which governs the dynamics of the mean value, and the other determines the evolution of the fluctuation operators. Next, we use the original quantum operators to represent its corresponding fluctuation operators for simplicity, e.g., $\hat{c}\Rightarrow\delta\hat{c} $, $\hat{P}_i\Rightarrow\delta\hat{P}_i $ and so on. Thus, the dynamics of the fluctuation operators can be explicitly given by
\begin{equation}
    \frac{d\mathbf {\hat u}(t)}{dt} = A\mathbf  {\hat u}(t) + \mathbf  {\hat n}_{\text{in}},
    \label{E4}
\end{equation}
where the column vector $\mathbf{\hat u}(t) = [\hat X_c(t), \hat P_c(t),\\\hat X_1(t),\hat X_2(t), \hat P_1(t), \hat P_2(t)]^\mathrm{T}$, the input noise $\mathbf{\hat n}_{\text{in}} = [\sqrt{2} \kappa \hat{X}_{c}^{\text{in}},\sqrt{2}\kappa \hat{P}_{c}^{\text{in}},0,0, \sqrt{2\gamma}f_{\text{in},1}, \sqrt{2\gamma}f_{\text{in},2}]^\mathrm{T}$, and
\begin{align}
\mathbf{A} = \begin{pmatrix}
-\kappa & \Delta' & 0 & 0 & 0 & 0 \\
-\Delta' & -\kappa & -G' & -G' & 0 & 0 \\
0 & 0 & 0 & V\sin \phi & \omega_{m} & V\cos \phi \\
0 & 0 & -V \sin \phi & 0 & V \cos \phi & \omega_{m} \\
-G' & 0 & -\omega_{m} & -V \cos \phi & -\gamma & V \sin \phi \\
-G' & 0 & -V \cos \phi & -\omega_{m} & -V \sin \phi &  -\gamma
\end{pmatrix},
\label{E5}
\end{align}
with $ G' = \sqrt{2}G=\sqrt{2}g\bar{c} $ and $\Delta' = \Delta + g (\bar{X}_1 + \bar{X}_2)$. In particular, in Eq.~(\ref{E5}), we have defined the optical quadrature operators $\hat{X}_c = \frac{\hat{c}^\dag + \hat{c}}{\sqrt{2}}$, $\hat{P}_c = i \frac{(\hat{c}^\dag - \hat{c})}{\sqrt{2}}$ and their corresponding input noise operators $\hat{X}_{c}^{\text{in}} = \frac{\hat{c}^{\dag \text{in}} + \hat{c}^{\text{in}}}{\sqrt{2}}$, $\hat{P}_{c}^{\text{in}} = i \frac{(\hat{c}^{\dag \text{in}} - \hat{c}^{\text{in}})}{\sqrt{2}}$.
 To solve Eq.~(\ref{E5}), we would like to change Eq.~(\ref{E5}) into the frequency domain by the Fourier transform 
 $ \hat o(\omega) = \int_{-\infty}^{+\infty} dt \, {e^{i\omega t}}\hat o(t)$
 for an arbitrary operator $\hat o(t)$, which yields
\begin{align}
\mathbf {\hat u}(\omega) = (-i\omega \mathbf{I} - A)^{-1} \mathbf{\hat n}_{\text{in}}(\omega),\label{E6}
\end{align}
where $\mathbf{I}$ is the identity matrix.

   A small external force will cause a slight equilibrium displacement of the oscillator, which in turn changes the cavity length, thereby altering the output phase of the optical cavity. Thus, it is possible to measure the quadrature phase of the cavity field output using a balanced homodyne detector. Here we use $\hat{P}^{o}_{\text{c}}$ and $\hat{X}^{o}_{\text{c}}$ to extract signals related to force and noise. The output fields are related to the input fields by input-output relations 
\begin{align}\label{E7}
  \hat{X}^{o}_{\text{c}} = \sqrt{2 \kappa} \, \hat{X}_{c} - \hat{X}^{in}_{\text{c}}, \quad
  \hat{P}^{o}_{\text{c}} = \sqrt{2 \kappa} \, \hat{P}_{c} - \hat{P}^{in}_{\text{c}}.
\end{align}
To detect the signal, we employ the homodyne detection. Thus, the output field will be mixed  a local strong oscillator with the phase $\theta$ through a 50:50 beam splitter. By adjusting the phase $\theta$, different quadrature components of the field can be measured. The photocurrent detected by the subtractor is proportional to the generalized quadrature of the output field as
 \begin{align} 
&\hat{P}^{o}_{\theta} = \cos\theta \hat{X}^{{o}}_{c} + \sin\theta \hat{P}^{{o}}_{c}\nonumber \\
&= A_1(\omega) \hat{X}_{c}^{{in}} + A_2(\omega)\hat {P}_{c}^{{in}}   + A_3(\omega) f_{\text{in,1}} + A_4(\omega) f_{{in,2}},\label{E8} 
 \end{align} 
 where all the coefficients $A_i$ are explicitly given in appendix \ref{appendix:derivation}.
Thus, the output symmetric power spectrum density can be given as  
	\begin{equation}\label{E9} 
		S(\omega)=\int d{\tilde{\omega}}\frac{e^{i(\omega+\tilde{\omega})t}}{4\pi}\langle \hat P^{{o}}_{\theta}(\omega)\hat P^{{o}}_{\theta}(\tilde{\omega})
		+ \hat{P}^{o}_{\theta}(\tilde{\omega})\hat{P}^{o}_{\theta}(\omega)\rangle.
	\end{equation}
Then one can easily derive the symmetric spectral density of the output field from Eq. (\ref{E9}) as
	\begin{equation}\label{11}
		S(\omega)=R_{m}^{\theta,\phi}(\omega)[S_{\rm{th}}^{\theta,\phi}(\omega)+N_{\rm{add}}^{\theta,\phi}(\omega)+S_{F_{ex}}],
	\end{equation}
where 	\begin{align}
		R_{m}^{\theta,\phi}(\omega)=&\vert{A_3}(\omega)+{A_4}(\omega)\vert ^2, \label{12}
  \end{align}
  \begin{align}
		S_{\rm{th}}^{\theta,\phi}(\omega)=&(\bar{n}+\frac{1}{2})\frac{ \vert {A_3}(\omega)\vert ^2+\vert {A_4}(\omega)\vert ^2}{\vert{A_3}(\omega)+{A_4}(\omega)\vert ^2},\label{13}
  \end{align}
  \begin{align}
  N_{\rm{add}}^{\theta,\phi}(\omega)=&\frac{1}{2}\frac{ \vert {A_1}(\omega)\vert ^2+\vert {A_2}(\omega)\vert ^2}{\vert{A_3}(\omega)+{A_4}(\omega)\vert ^2},\label{14}
	\end{align}
	represent the response to external signals, the non-dimensional thermal noise of the two mechanical oscillators, and the additional noise of weak force field sensing, respectively,
and $S_{F_{ex}}$ is the spectral density of the signal generated by the weak force to be measured. From Eq. (\ref{12}), one can see that $R_{m}^{\theta,\phi}(\omega)>1$ indicates the signal amplification.  The additional noise $N_{\rm{add}}^{\theta,\phi}(\omega)$ can be divided into back-action noise and shot noise. When the optomechanical coupling strength reaches its optimal value, the noise achieves its minimum.
 The smaller the additional noise $N_{\rm{add}}^{\theta,\phi}(\omega)$ and the thermal noise $S_{\rm{th}}^{\theta,\phi}(\omega)$ of the system are, the easier it is to detect the signal of the external weak force. Therefore, reducing the noise will improve the sensitivity of weak force sensing. In this sense, we won't consider the specific expression of the detected weak force signal spectrum but focus on reducing the noise and enhancing the response.


\section{THE NOISE SUPPRESSION PERFORMANCE UNDER SYNTHETIC MAGNETIC CONDITIONS}
  \label{sec: results}

In this section, We demonstrate the synthetic magnetism induced by tuning the coupling strength and phase parameters between mechanical oscillators. We then significantly suppress additional noise and enhance the mechanical response by leveraging synthetic magnetism to break dark modes. This process can even result in mode splitting and frequency shifts, expanding the detection bandwidth. We also explore how, with a fixed coupling strength between the mechanical resonators and varying phases, adjusting the system's effective optomechanical coupling parameters and dissipation coefficient of the cavity mode can reduce the additional noise. Finally, we examine the influence of synthetic magnetism on thermal noise. For simplicity, we take the homodyne phase angle $\theta = \frac{\pi}{2}$.
\subsection{Mechanical response and the Added noise}
    \begin{table}[t]
        \centering
        \caption{System parameters used in all cases}
        \begin{tabular}{l|c|l}\hline\hline
            mechanical Resonator frequency  & $\omega_m$            & $2\pi\times3.6$ MHz  \\
            Temperature    & $T$               & $0.077$ K    \\
            Cavity decay rate    & $\kappa$            & $2\pi\times360$ kHz  \\
            Resonator decay rate & $\gamma$            & $2\pi\times36$ Hz     \\
            Cavity-field effective detuning   & $\Delta^{\prime}$     & $0$                  \\\hline\hline
        \end{tabular}
        \label{tab:parameters}
    \end{table} 
    \begin{figure}[b]
        \centering
        \includegraphics[width=8cm]{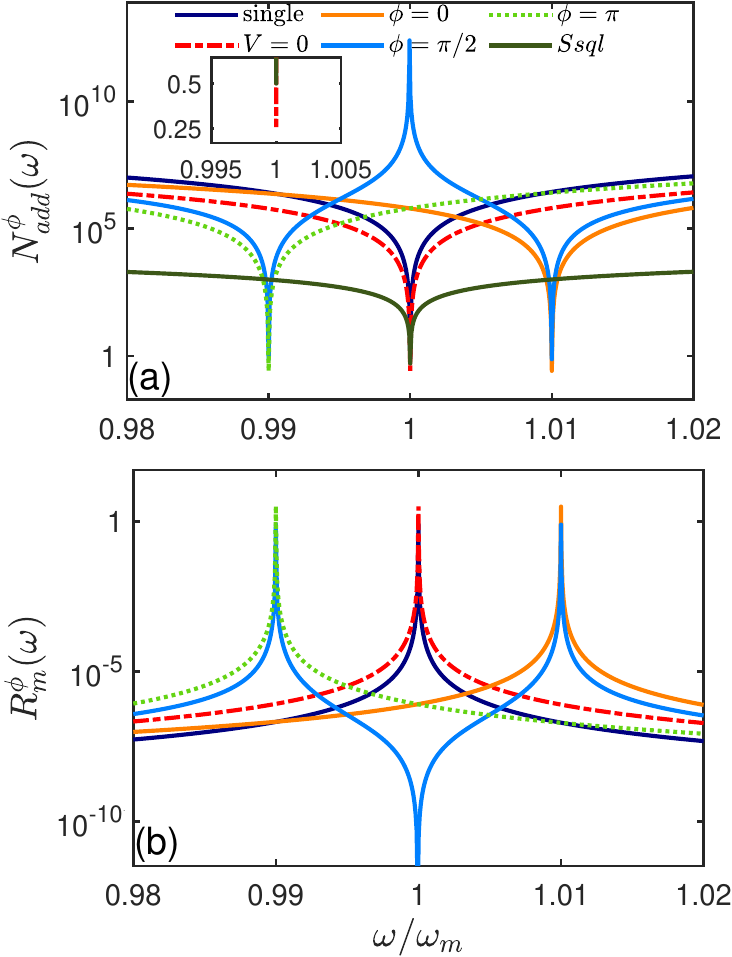}
        \caption{The dimensionless additional noise power spectral density $N_{\text{add}}^{\phi}(\omega)$ (a) and the mechanical response $R_{m}^{\phi}(\omega)$ (b) as functions of the normalized frequency $\omega/\omega_m$ for various coupling phases $\phi$. {The system parameters are based on those provided in Table \ref{tab:parameters} with $V = 0.01 \omega_m$ and $G' = 4.5 \times 10^{-3} \, \omega_m$.}}
        \label{fig:2}
    \end{figure}
\begin{figure}[b]
        \centering
        \includegraphics[width=\linewidth]{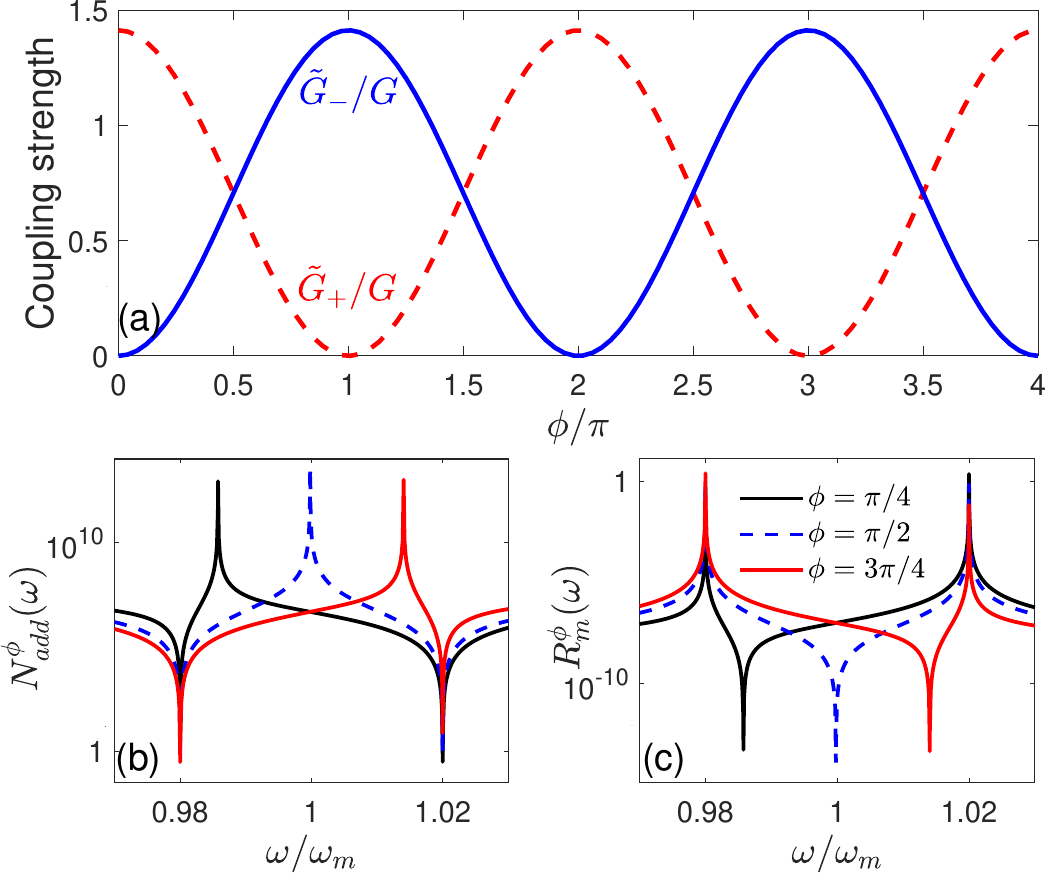}
        \caption{(a) The effective coupling strengths $\tilde{G}_{\pm}$ as functions of $\phi$. The dimensionless additional noise power spectral density $N_{\rm{add}}^{\phi}(\omega)$ (b) and the mechanical response $R_{m}^{\phi}(\omega)$ (c) as functions of the normalized frequency $\omega/\omega_m$ for various coupling phases $\phi$. { Here the parameters are consistent with Fig. \ref{fig:2} and $V = 0.02  \omega_m$}.
}        \label{fig:3}
    \end{figure}
     In Fig. \ref{fig:2} (a) and (b), we plot $N_{\rm{add}}^{\phi}(\omega)$ and $R_{m}^{\phi}(\omega)$ as functions of the detection frequency.  Whether the two mechanical oscillators are coupled or not can be controlled by the parameter $V$. When \( V = 0 \), the effective detection frequency \( \omega_{\mathrm{eff}} = \omega_m \) matches the resonance frequency of the oscillators, similar to a single oscillator.
 However, the distinction is that the dual-oscillator system exhibits reduced additional noise, achieving a minimum added noise as low as $0.25$, thus surpassing the SQL $0.5$.
    Furthermore, as shown in Fig. \ref{fig:2} (b), the two ports  enable coherent amplification during the signal conversion process, resulting in a mechanical response greater than 1
 \cite{PhysRevA.99.063811}. 

    {When $ V = 0.01\omega_m $}, adjusting the mechanical coupling  $\phi$ reveals that the effective detection frequency undergoes a shift. Specifically, for $\phi = 0$, {the effective frequency emerges in the high-frequency region at $\omega_{\mathrm{eff}} = 1.01\omega_m$}, whereas for $\phi = \pi$, {it appears in the low-frequency region at $\omega_{\mathrm{eff}} = 0.99 \omega_m$}. Furthermore, the effective frequency shift relative to $\omega_m$ exhibits a symmetrical behavior under the same coupling strength.
Additionally, we observe that when $\phi = n\pi$, a single split mode is present, whereas, for $\phi \neq n\pi$, the effective detection frequency manifests at two symmetric high and low frequencies relative to $\omega_m$, {i.e., $\omega_{\text{\textup{eff}}} = 0.99 \omega_{m}$ and $\omega_{\text{\textup{eff}}} = 1.01 \omega_{m}$ at $\phi = \frac{\pi}{2}$}. However, the enhancements in mechanical response and noise reduction are slightly diminished compared to the absence of mode splitting, as the activated phonons are distributed between the two channels. Overall, with phase coupling in the system, the added noise $N_{\text{add}}^{\phi}(\omega)$ at the effective detection frequency closely resembles that observed when the two oscillators are uncoupled,and a similar trend is observed in the mechanical
response. 
\begin{figure*}[t]
\centering
\includegraphics[width=0.31\textwidth]{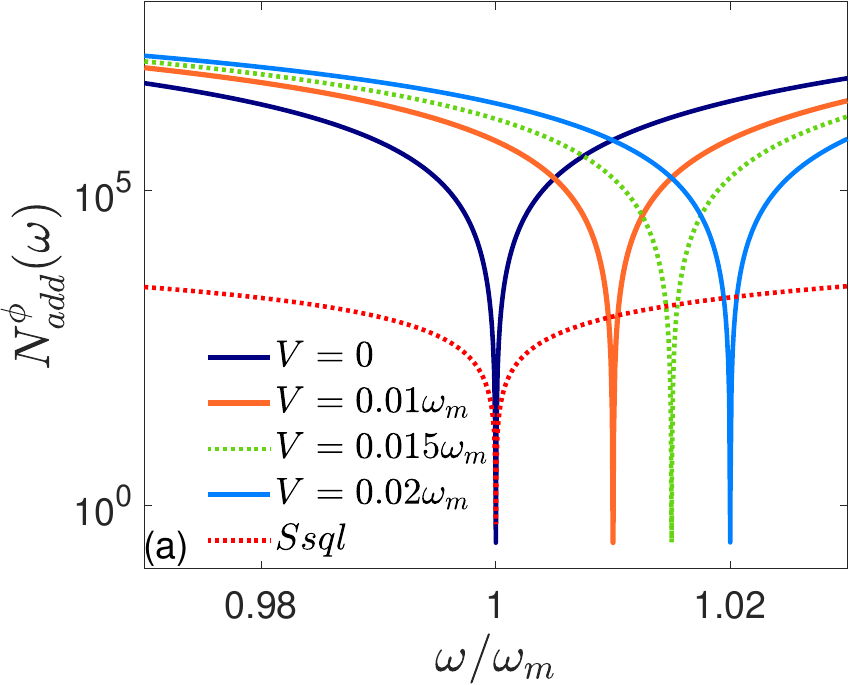}  \label{fig:nadd211.eps-58144(1)}
\hspace{0.001\textwidth}
\includegraphics[width=0.31\textwidth]{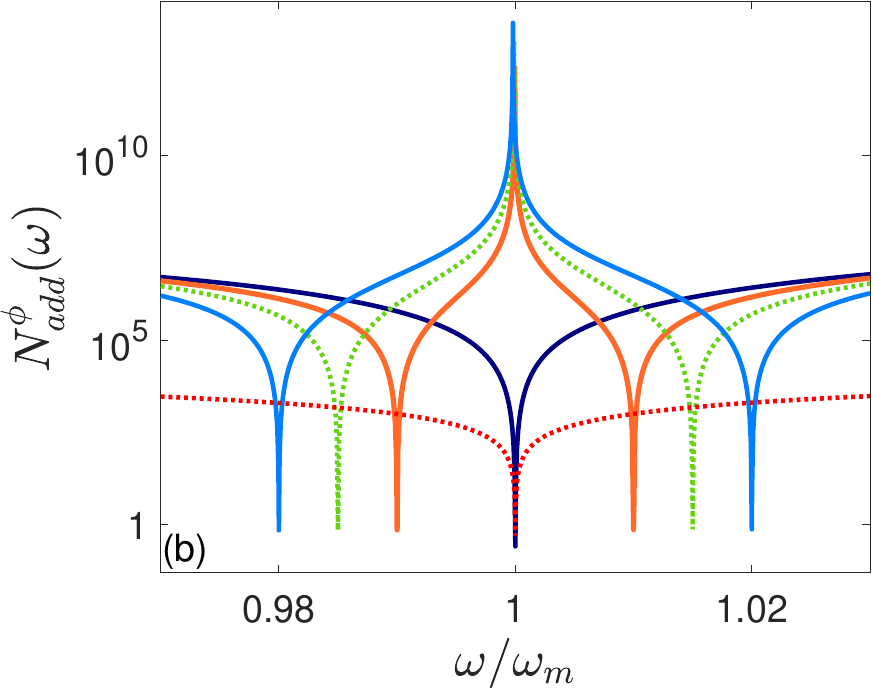} \label{fig:nadd222.eps-73023(1)} 
\hspace{0.001\textwidth}
\includegraphics[width=0.31\textwidth]{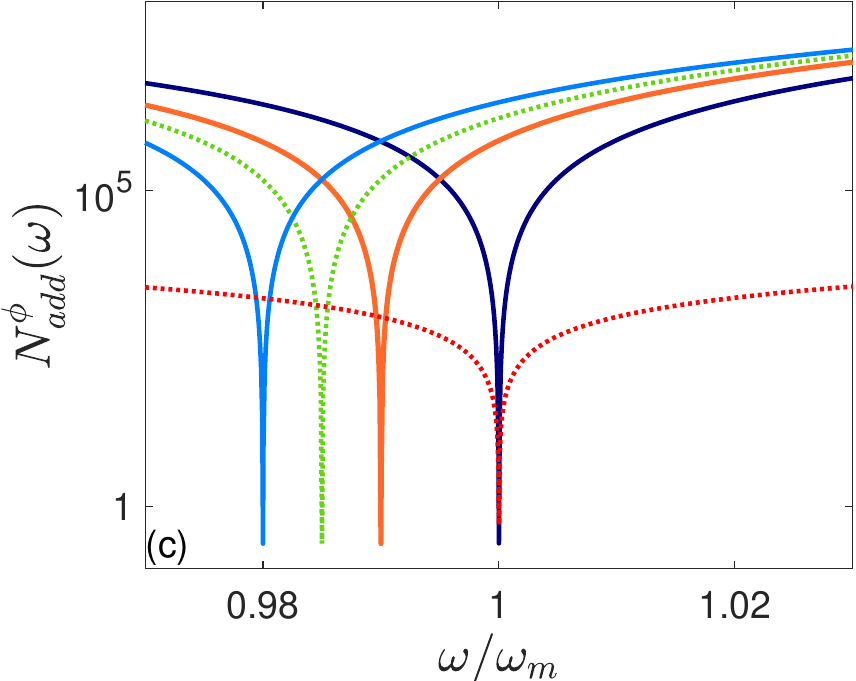} 
\caption{The dimensionless additional noise power spectral density $N_{\text{add}}^{\phi}$ as the function of $\omega/\omega_m$ for different coupling phases $\phi = 0$ (a), $\phi = \frac{\pi}{2}$ (b), and $\phi = \pi$ (c) for different coupling strength $V$, respectively. The parameters are taken from Table \ref{tab:parameters}, and $G' = 4.5 \times 10^{-3}
 \omega_m$.}
\label{fig:nadd231(1)}
\end{figure*}

 {To provide a more comprehensive understanding of this phenomenon, we diagonalize the Hamiltonian and express it in terms of the bosonic creation and annihilation operators \( \hat{b}_j^\dagger \) and \( \hat{b}_j \). The linearized optomechanical Hamiltonian, as derived from Eq.~(\ref{E1}), is given by
\begin{align}
&\hat{H} = \hbar \Delta' \hat{c}^\dagger \hat{c} +\hbar \omega_m \sum_{i=1}^{2} \hat b_i^\dagger\hat b_i \ + \hbar G \sum_{i=1}^{2} (\hat c\hat b_i^\dagger +\hat c^\dagger\hat b_i ) \nonumber \\
&\quad +\hbar V ( e^{i\phi}\hat b_1^\dagger\hat b_2 + e^{-i\phi}\hat b_1 \hat b_2^\dagger )  -x_{\text{ZPF}} F(t)\displaystyle\sum\limits_{i=1,2}(\hat{b}_i + \hat{b}_i^\dagger).
\label{E15}
\end{align}}
 The synthetic magnetism results in path interference between two excitation transfer channels. The two superposed vibrational modes closely linked to synthetic magnetism are given as 
\begin{eqnarray}
 \tilde{\mathcal\hat{B}}_+ =\frac{1}{\sqrt{2}}\hat b_1 + e^{i\phi} \frac{1}{\sqrt{2}}\hat b_2, \notag \\
  \tilde{\mathcal\hat {B}}_- = -\frac{1}{\sqrt{2}}  e^{-i\phi} \hat b_1 + \frac{1}{\sqrt{2}}\hat b_2.\label{E16}
\end{eqnarray}
Thus the Hamiltonian in  Eq. (\ref{E15}) can be  rewritten as
  \begin{align}
 \hat{H} &= \hbar \Delta' \hat{c}^\dagger \hat{c} 
  + \hbar \tilde{\omega}_{+} \tilde{\mathcal\hat{B}}_{+}^\dagger \tilde{\mathcal\hat {B}}_{+} 
  +\hbar \tilde{ \omega}_{-} \tilde{\mathcal\hat{B}}_{-}^\dagger \tilde{\mathcal\hat {B}}_{-} \nonumber \\
  &\quad +\hbar \tilde{G}_{+}^{*}\hat c \tilde{\mathcal\hat {B}}_{+}^\dagger + \hbar \tilde{G}_{+} \hat c^\dagger  \tilde{\mathcal\hat {B}}_{+} +\hbar \tilde{G}_{-}^{*}\hat c \tilde{\mathcal\hat {B}}_{-}^\dagger +\hbar \tilde{G}_{-}\hat  c^\dagger \tilde{\mathcal\hat {B}}_{-} \nonumber \\
   &\quad +\frac{1}{\sqrt{2}} x_{\text{ZPF}} F(t)((1+e^{-i\phi})\tilde{\mathcal\hat {B}}_{+}+(1+e^{i\phi})\tilde{\mathcal\hat {B}}_{+}^\dagger)\nonumber\\&\quad+(1-e^{i\phi})\tilde{\mathcal\hat {B}}_{-}+(1-e^{-i\phi})\tilde{\mathcal\hat {B}}_{-}^\dagger),\label{E17}
\end{align}
which is similar to the center-of-mass and relative motions in two coupled classical oscillators. The effective coupling strength and frequency of the two mechanical oscillators are redefined as \cite{PhysRevLett.129.063602} 
\begin{eqnarray}
\tilde{\omega}_{\pm}=\omega_{m} \pm V,\quad
\tilde{G}_{\pm} = \frac{1}{\sqrt{2}} G \left(1 \pm e^{\mp i \phi}\right).\label{E18}
\end{eqnarray}
As shown in Eq.~(\ref{E17}), when \(\phi = n\pi\) with integer \(n\), one of the hybrid mechanical modes (the dark mode) decouples from the external force signal and the cavity mode. This condition corresponds to \(\phi = 0\) (orange lines) and \(\phi = \pi\) (green lines) in Fig. \ref{fig:2}, where the effective probe frequency appears as a single mode without mode splitting.

In Fig. \ref{fig:3} (a), we plot $\tilde{G}_\pm$ as the function of $\phi$. When $\phi = n\pi$, only one effective frequency exists.  Due to coherent interference, specific degenerate vibrational modes can be decoupled from the cavity field, corresponding to the unbroken dark mode (DMU).
Specifically, when $n$ is odd, $\tilde{G}_+ = 0$ (the red  line in Fig. \ref{fig:3} (a)). Accordingly, \( \tilde{B}_+ \) corresponds to the dark mode, whereas \( \tilde{B}_- \) represents the bright mode. {Thus the effective detection frequency appears in the low-frequency range (the green line in Fig.~\ref{fig:2})}. Conversely, when $n$ is even, $\tilde{G}_- = 0$ (the blue  line in Fig. \ref{fig:3} (a)), $\tilde{B}_-$ corresponds to the dark mode and $ \tilde{B}_+ $ represents the bright mode.{ The effective detection frequency is associated with the high-frequency range ( the orange line in Fig.~\ref{fig:2}).} At $\phi = \frac{\pi}{2}$, corresponding to the dark mode being broken (DMB), an unexpected coupling occurs between the vibrational and optical modes, leading to the splitting of the mechanical modes—this splitting results from synthetic magnetism. 

In Fig. \ref{fig:3} (b) and (c), we consider two coupling phases symmetric about $\phi = \frac{\pi}{2}$, specifically $\phi = \frac{\pi}{4}$ and $\phi = \frac{3\pi}{4}$, we also plot the relations of $N_{add}^{\phi}$ and $R_{m}^{\phi}$ as functions of $\omega/\omega_m$ . We find that at $\phi = \frac{\pi}{4}$ and $\phi = \frac{3\pi}{4}$, { added noise reaches its minimum values around $0.98\omega_m$ and $1.02\omega_m$}. However, at the lower frequency $\omega_{\mathrm{eff}} = 0.98\omega_m$, the added noise for  $\phi = \frac{3\pi}{4}$ is lower compared to $\phi = \frac{\pi}{4}$, and the corresponding mechanical response is higher for $\phi = \frac{3\pi}{4}$. {Similarly, when $\phi = \frac{3\pi}{4}$, mechanical response at $\omega_{\mathrm{eff}} = 0.98\omega_m$ is 3.6, while at $\omega_{\mathrm{eff}} = 1.02\omega_m$, mechanical response is 0.04,} indicating that the sensitivity at $\phi = \frac{3\pi}{4}$ is significantly higher at lower frequencies.  Similarly, the sensitivity at $\phi = \frac{\pi}{4}$ is significantly higher at higher frequencies. According to Fig. \ref{fig:3}, when $\phi = \frac{3\pi}{4}$, the coupling strength $\tilde{G}_-$ is greater than $\tilde{G}_+$, when $\phi = \frac{\pi}{4}$, the coupling strength $\tilde{G}_-$ is less than $\tilde{G}_+$. For $\phi = \frac{\pi}{2}$, the detection efficiencies at the two detection frequencies are nearly identical.
  Based on  Eq. (\ref{E18}), it can be concluded that the larger the ratio $\tilde{G}_-/\tilde{G}_+$, the greater quantum channel coupling the vibrational mode at $\omega_-$ (which corresponds to the lower effective frequency) to the cavity field. On the contrary, if the larger the ratio $\tilde{G}_+/\tilde{G}_-$, the greater the quantum channel coupling the vibrational mode at $\omega_+$.
Therefore, when $0 \leq \phi < \frac{\pi}{2}$ and $\frac{3\pi}{2} < \phi \leq 2\pi$, the detection effect at high frequencies is superior to that at low frequencies; when $\frac{\pi}{2} < \phi < \pi$, and $ \pi < \phi < \frac{3\pi}{2}$,the detection effect at low frequencies is superior to that at high frequencies. {When an in-phase force acts on the two modes, amplification occurs in only one mode. This asymmetry originates from the modulation phase in the coupling circuit, driven by intensity differences between two quantum interference pathways. These differences significantly affect the mechanical response, as varying \( \phi \) modifies the relative coupling strengths \( \tilde{G}^- \) and \( \tilde{G}^+ \), resulting in distinct sensitivities at low and high frequencies. Similar behavior has been observed in nanomechanical coupled resonators, where cantilevers exhibit opposing stochastic responses due to variations in stiffness and structural dissipation \cite{doi:10.1021/nl902350b}.
} 

 Next, to more clearly explore the relationship between the coupling strength $V$, the coupling  $\phi$, and the added noise spectral density,
we plot $N_{\text{add}}^{\phi}$ versus the detection frequency in Fig. \ref{fig:nadd231(1)}.
 Regardless of the coupling phase $\phi$, the two coupled oscillators' motion in-phase and out-of-phase modes leads to changes in the system's natural frequency according to Eq. (\ref{E18}). This frequency shift entirely depends on the coupling strength of the mechanical oscillators $V$. In Fig . \ref{fig:nadd211.eps-58144(1)} (a) and Fig. \ref{fig:nadd231(1)} (c), $\phi$ is set to be $0$ and $\pi$, respectively, corresponding to the even and odd breaking of the dark modes, with the effective detection frequencies distributed on the higher and lower sides of $\omega_m$. Only a frequency shift occurs without mode splitting, which is consistent with the conclusion we draw in Fig. \ref{fig:2}. Introducing an additional oscillator allows the system to surpass the SQL at the effective frequency $\omega_{\text{eff}}$, with the coupling strength only affecting the frequency shift, regardless of the presence of coupling. In Fig. \ref{fig:nadd222.eps-73023(1)} (b), with \(\phi = \frac{\pi}{2}\), by breaking the dark mode and adjusting the coupling strength, effectively increases the detection bandwidth. Also the added noise at each probe frequency can break SQL. Thus when breaking dark modes through synthetic damping, we can achieve simultaneous detection of both high and low-frequency signals. The coupling strength $V$ can be manipulated to modify the distance from the amplification frequency, while the coupling phase $ \phi $ can direct the effective detection frequency.

\begin{figure}[t]
        \centering
        	\centering\includegraphics[width=\linewidth]{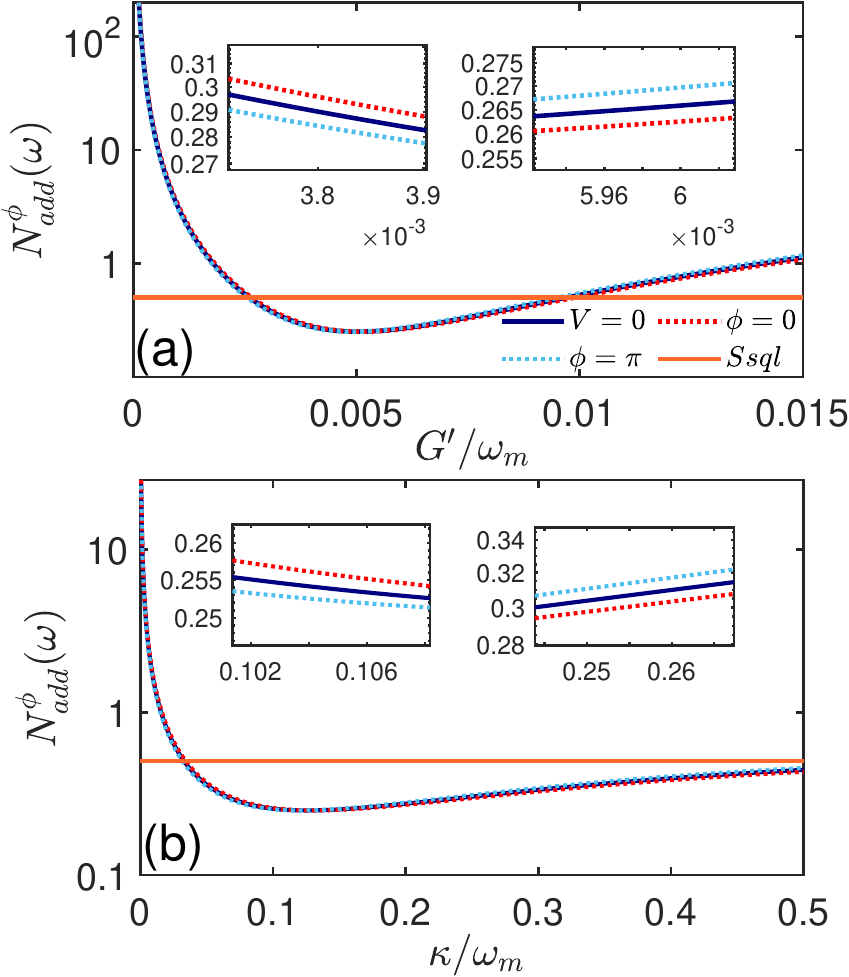}
        \caption{(a) The dimensionless additional noise power spectral density $N_{\text{add}}^{\phi}$ at $\omega_\text{eff}$ as the function of $G'/\omega_m$
 for various coupling $\phi$, with $\kappa = 0.1 \omega_m$.(b) The dimensionless added noise power spectral density $N_{\text{add}}^{\phi}$ at $\omega_\text{eff}$ as the function of $\kappa/\omega_m$ for different coupling phases, with $G' = 4.5 \times 10^{-3}  \omega_m$. The parameters for both (a) and (b) are taken from Table \ref{tab:parameters}, and {the coupling strength of the two oscillator $V = 0.02 \, \omega_m$.}
}        \label{fig:abc}
    \end{figure}

To investigate the relationship between added noise and other tunable parameters, we plot Fig. \ref{fig:abc}. The added noise values were obtained at the corresponding effective frequencies for various coupling phases. { For $\phi = 0$ , $\omega_{\text{eff}} = 1.02\omega_m$, the effective frequency without coupling is $\omega_{\text{eff}} = \omega_m$, while for $\phi = \pi$, $\omega_{\text{eff}} = 0.98\omega_m$.} Back-action noise dominates at low frequencies \cite{clerk2010introduction}, while shot noise dominates at high frequencies. In Fig. \ref{fig:abc} (a), as the coupling strength increases, the added noise first decreases and then increases, reaching a minimum at $G' = 5 \times 10^{-3} \omega_m$, surpassing the SQL and indicating  our scheme can realize highly sensitive weak force detection without the need of ultra-strong coupling or deep-strong coupling mechanism, which reduces the difficulty of experimental.  This behavior arises from the competitive interplay between back-action noise and shot noise. At small $G'$, the primary reduction occurs in photon shot noise due to lower laser driving intensities. The additional noise at low frequencies is lower than that at high frequencies, as the spring effect induces a `softer mode'
 at low frequencies. This softer mode is advantageous for suppressing shot noise, offering valuable insights for reducing quantum noise in LIGO.
As $G'$ increases, back-action noise induced by radiation pressure also rises and high frequencies is greater than that at low frequencies.  
In Fig. \ref{fig:abc} (b), a similar trend is observed: the added noise decreases initially and then increases as the coupling strength rises. When the dissipation is minimal, extracting information from the mechanical oscillators becomes challenging, resulting in increased noise. As the dissipation gradually increases, information retrieval becomes more efficient, leading to a reduction in added noise. However, with further increases in dissipation, quantum noise from the cavity field begins to dominate, causing the noise to rise again.
\subsection{{Thermal mechanical noise}}
\label{subsec:thermal noise}
\begin{figure}[b]    
        	\centering\includegraphics[width=\linewidth]{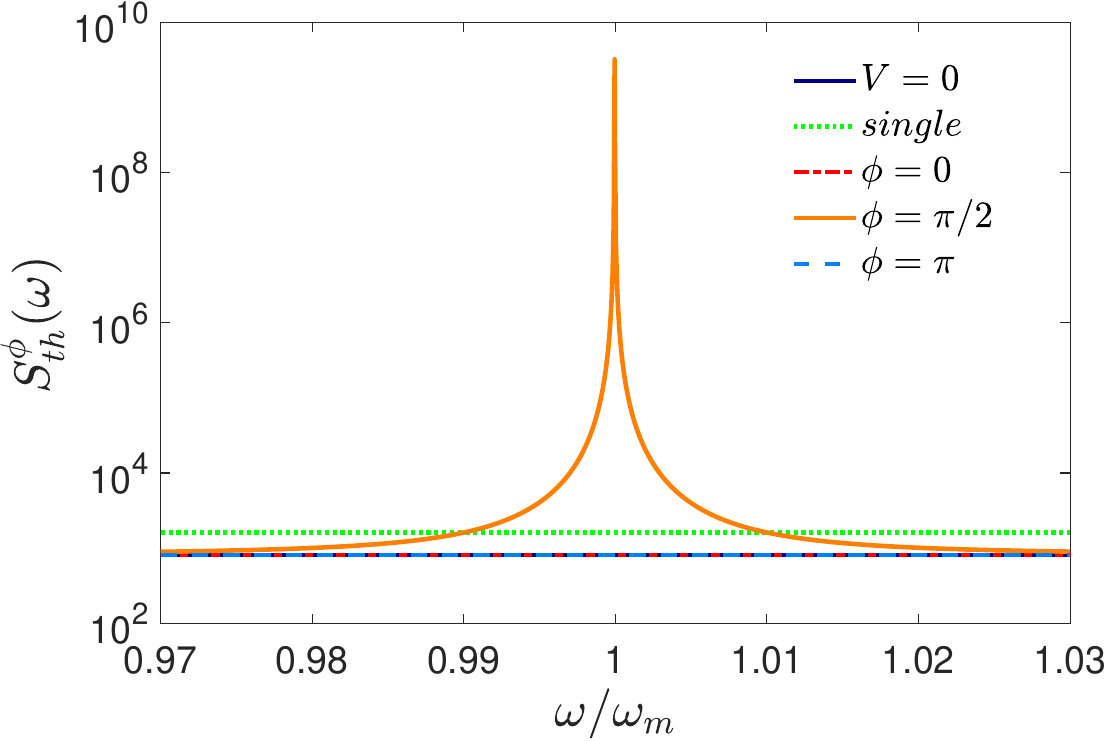}
        \caption{The dimensionless thermal noise power spectral density $S_{\rm{th}}^{\phi}(\omega)$ at the $\omega_\text{eff}$ as the function of $\omega/\omega_m$ for various coupling phases, {with $V = 0.01 \omega_m$ and $G' = 4.5 \times 10^{-3}\omega_m$.} Other parameters are taken from Table \ref{tab:parameters}.
        \label{fig:7}
}
    \end{figure}

   Fig. \ref{fig:7}  displays the thermal noise $S_{\rm{th}}^{\phi}(\omega)$ as the function of normalized frequency $\omega/\omega_m$.It shows that increasing the mechanical oscillators does not increase the effective thermal noise. On the contrary, when there is no coupling between the two oscillators, the thermal noise is reduced by half compared to the standard system subject to the identical oscillator parameters. This is because the quantum correlations between different oscillators can be neglected. When the oscillators are coupled with the phase $0$ or $\pi$, the dark mode remains unbroken. This indicates that the optomechanical cooling channel cannot extract the thermal phonon number hidden in the dark mode, resulting in lower thermal noise than a single oscillator. When the phase is  $\phi = \frac{\pi}{2}$ , the dark mode is broken, allowing the coherent cooling channel to extract the thermal phonon number hidden in the dark mode.
{Consequently, at $V = 0.01 \omega_m$, specifically at effective frequencies of $0.99 \omega_m$ and $1.01 \omega_m$}, the thermal noise matches that of a single oscillator, with neither suppression nor enhancement. Near $\omega = \omega_m$, $S_{\rm{th}}^{\phi}(\omega)$ shows a peak. Fortunately, this frequency is not utilized. Additionally, when the dark mode is broken, the thermal noise distribution is nearly symmetric around $\omega_m$, which coincides with our previous discussion that both thermal noise and the signal enter the optomechanical system through the same channel, leading to mode splitting.


\maketitle
\section{EXPERIMENTAL IMPLEMENTATIONS}
\label{sec:experiments}

	{In practical applications, achieving a synthetic magnetic field and maintaining robust mechanical modes with a quality factor ($Q \approx 10^{5}$) often presents significant experimental challenges. However, the approach we propose can be realized through a photonic crystal optomechanical system \cite{eichenfield2009picogram, xia2020opto}. We identify the crucial role of silicon nitride (SiN) thin films in facilitating the coupling and modulation of hybrid modes \cite{PhysRevLett.115.017202}, particularly in addressing dissipative effects associated with near-intrinsic modes. Based on this, the coupling between two resonators can be effectively controlled via optothermal modulation of the SiN surface substrate interactions. By generating synthetic dimensions between the two optomechanical cavities, a synthetic magnetic effect can be achieved, where radiation pressure coupling forms a two-dimensional lattice between the optical and mechanical modes. In this lattice, Photons and phonons hop with different rates, respectively. The two cavities are driven by phase-locked laser pumps, with the phase difference controlling the generation of the synthetic magnetic field. Synthetic flux is mediated by the central silicon beam connecting the cavities, which in turn facilitates photon-phonon hopping. The phase difference can be stabilized by a fiber optic phase modulator, enabling the realization of a stable synthetic magnetic field \cite{fang2017generalized, Habraken_2012}. Synthetic magnetic effects can be realized across various experimental platforms. For instance, in a system composed of two mechanical resonators and two microwave cavity modes, a quadruple modulation signal precisely controls the interactions, thereby inducing a synthetic gauge field \cite{doi:10.1126/science.abf5389}. In another setup, where a micromechanical resonator is coupled with a superconducting circuit, cavity frequency modulation via capacitance simulates radiation pressure effects, while microwave pump signals induce two-mode squeezing and entanglement, generating the synthetic magnetic field \cite{ockeloen2018stabilized}. Additionally, by coupling superconducting qubits to surface acoustic wave (SAW) cavities, Floquet engineering can be realized in a three-cavity loop, inducing a synthetic gauge field \cite{Wang_2020}.}
\section{Conclusion}
    \label{sec:conclusion}
In summary, we have explored the enhancement of weak force sensing using synthetic magnetism in a two-mechanical mode optomechanical system. Our results demonstrate that adjusting the coupling strength between the two oscillators can shift the effective detection frequency. Moreover, by tuning the coupling phase $\phi$, synthetic magnetism enables flexible switching between broken and unbroken dark modes. When $\phi = n\pi$, the dark mode is unbroken, allowing control over the direction of the effective detection frequency shift, thus enabling switching between high and low frequencies in multimode quantum devices. Conversely, when $\phi \neq n\pi$, the dark mode remains broken, leading to mode splitting. This results in a uniform broadband frequency response at the resonance frequency, significantly overcoming the limitation of the narrow frequency window caused by mechanical resonance. This is crucial for mechanical sensing applications requiring broadband detection. Therefore, by adjusting $\phi$, synthetic magnetism provides excellent frequency tunability for mechanical sensing applications.

Our scheme also effectively suppresses additional noise, breaking the SQL at the effective detection frequency. By utilizing the quantum correlations between the two oscillators and selecting appropriate optomechanical coupling strength and cavity field dissipation, additional noise can be significantly reduced without weakening the signal. Besides, in our scheme, under the broken dark mode condition, thermal noise is reduced. Under the unbroken condition of the dark mode, thermal noise remains unaffected at the effective frequency, meaning our scheme can achieve a higher signal-to-noise ratio, thereby improving signal sensitivity and resolution. Our approach provides a promising platform for optomechanical systems and quantum weak force detection.


\section*{Acknowledgements}
We thank Shi-wen He and Ye-ting Yan and Yu-qiang Liu for their helpful discussions. This work was supported by the National Natural Science Foundation of China under Grant No. 12175029. 

\appendix

\renewcommand{\thefigure}{A\arabic{figure}}
\setcounter{figure}{0}
 
\section{The Detailed Coefficient of the Output Quadrature Components}
\label{appendix:derivation}

 The generalized quadrature operatures in Eq. (\ref{E6}) of the cavity field mode is given by
\begin{align} 
  \hat {X}_{\text{c}} =&\, {k_1}(\omega)\hat {X}_{c}^{\text{in}} + {k_2}(\omega)\hat {P}_{c}^{\text{in}} + {k_3}(\omega) f_{\text{in,1}} + {k_4}(\omega) f_{\text{in,2}},\notag \\
 \hat {P}_{\text{c}} =&\, {k_5}(\omega) \hat {X}_{c}^{\text{in}} + {k_6}(\omega)\hat {P}_{c}^{\text{in}} + {k_7}(\omega) f_{\text{in,1}} + {k_8}(\omega) f_{\text{in,2}}.
\end{align}

The coefficients of the generalized amplitude and phase quadrature operators of the cavity field are are given as
\begin{widetext}
\begin{align}
k_1(\omega) &= \frac{\sqrt{2} \sqrt{\kappa} (\kappa - i \omega) (  \gamma^2 V^2 \cos 2 \phi - e_6 )}{-e_1 e_2 - e_3 + e_4 - e_5}, 
k_2(\omega) = \frac{k_1(\omega) \Delta}{\kappa - i \omega}, \notag \\
k_3(\omega) &= \frac{2 \sqrt{2} \Delta g \sqrt{\gamma} (-(V^2 + \omega (i \gamma + \omega)) \omega_m + \omega_m^3 - V (\gamma - 2 i \omega) \omega_m \sin\phi + V \cos\phi (V^2 - i \gamma \omega - \omega^2 - \omega_m^2 + \gamma V \sin\phi))}{-e_1 e_2 - e_3 + e_4 + e_5}, \notag \\
k_4(\omega) &= k_3(\omega), 
k_5(\omega) = \frac{\sqrt{2} \sqrt{\kappa} (4 g^2 \omega_m (V^2 + i \gamma \omega + \omega^2 - \omega_m^2) + \Delta  e_6  + e_6 \cos\phi - \Delta \gamma^2 V \cos2 \phi )}{-e_1 e_2 - e_3 + e_4 + e_5}, \notag \\
k_6(\omega) &= k_1(\omega),k_7(\omega) = \frac{k_3(\omega) \Delta}{\kappa - i \omega}, k_8(\omega) = k_7(\omega).
\end{align}
with
\begin{align}
e_1 &= \Delta^2 + (\kappa - i \omega)^2, \notag \\
e_2 &= \gamma^2 (V^2 - 2 \omega^2) + 2 (V^2 - \omega^2)^2 + 4 i \gamma \omega (-V^2 + \omega^2), \notag \\
e_3 &= 4 \Delta g^2 (V^2 + \omega (i \gamma + \omega)) \omega_m, \notag \\
e_4 &= 4 e_1 (V^2 + \omega (i \gamma + \omega)) \omega_m^2 - 2 e_1 \omega_m^4, \notag \\
e_5 &= 4 \Delta g^2 V (V^2 - i \gamma \omega - \omega^2 - \omega_m^2) \cos\alpha + \gamma^2 V^2 e_1 \cos 2\phi, \notag \\
e_6 &= \gamma^2 (V^2 - 2 \omega^2) + 2 (V - \omega - \omega_m)(V + \omega - \omega_m)(V - \omega + \omega_m)(V + \omega + \omega_m) - 4 i \gamma \omega (V^2 - \omega^2 + \omega_m^2).
\end{align}
\end{widetext}
According to the input-output relations in Eq. (\ref{E7}), the generalized operators of the output field are given by
\begin{align}  
  \hat{X}^{\text{out}}_{\text{c}} =&\, \tilde{k}_1(\omega) \hat {X}_{c}^{\text{in}} + \tilde{k}_2(\omega)\hat {P}_{c}^{\text{in}} +\tilde{k}_3(\omega) f_{\text{in,1}} + \tilde{k}_4(\omega) f_{\text{in,2}},\notag \\
 \hat {P}^{\text{out}}_{\text{c}} =&\, \tilde{k}_5(\omega)\hat {X}_{c}^{\text{in}} + \tilde{k}_6(\omega)\hat {P}_{c}^{\text{in}} + \tilde{k}_7(\omega) f_{\text{in,1}} + \tilde{k}_8(\omega) f_{\text{in,2}}.
\end{align}
with
\begin{align}
	\tilde{k}_i &= \sqrt{2\kappa} ( k_i - \frac{1}{\sqrt{2\kappa}} ), 
	\quad  i = 1, 5, \notag \\
	\tilde{k}_i &= \sqrt{2\kappa} k_i,  \quad  i = 2,3,4,6,7,8.
\end{align}
Thus, the coefficients of the input noise operators in  Eq. (\ref{E8}) are given by
\begin{align} 
A_i = \cos \theta \,\tilde{k}_i  + \sin \theta \, \tilde{k}_{i+4} ,  \quad  i = 1, 2, 3, 4.
\end{align}

\bibliography{main}

\end{document}